\documentclass[aps,prl,twocolumn,amsfonts,amssymb,amsmath,floats,floatfix,showpacs,preprintnumbers,superscriptaddress,unsortedaddress]{revtex4} 

\usepackage{graphicx}
\usepackage{dcolumn}
\usepackage{bm}
\usepackage{color}

\usepackage{amsmath}

\newcommand{\av}[1]{ \langle #1 \rangle}
\newcommand{\figdir}{./}

\begin{document}

\title{A spectral function tour of electron-phonon
  coupling outside the Migdal limit}

\author{C.\,N. Veenstra}
\affiliation{Department of Physics {\rm {\&}} Astronomy, University of British Columbia, Vancouver, British Columbia V6T\,1Z1, Canada}
\author{G.\,L. Goodvin}
\affiliation{Department of Physics {\rm {\&}} Astronomy, University of British Columbia, Vancouver, British Columbia V6T\,1Z1, Canada}
\author{M. Berciu}
\affiliation{Department of Physics {\rm {\&}} Astronomy, University of British Columbia, Vancouver, British Columbia V6T\,1Z1, Canada}
\affiliation{Quantum Matter Institute, University of British Columbia, Vancouver, British Columbia V6T\,1Z4, Canada}
\author{A. Damascelli}
\email{damascelli@physics.ubc.ca}
\affiliation{Department of Physics {\rm {\&}} Astronomy, University of British Columbia, Vancouver, British Columbia V6T\,1Z1, Canada}
\affiliation{Quantum Matter Institute, University of British Columbia, Vancouver, British Columbia V6T\,1Z4, Canada}
\date{\today}

\begin{abstract}
We simulate spectral functions for electron-phonon coupling in a
filled band system - far from the asymptotic limit often
assumed where the phonon energy is very small compared to
the Fermi energy in a parabolic band and the Migdal theorem 
predicting $(1+\lambda)$ quasiparticle renormalizations is valid.
These spectral functions are examined over a wide range of parameter
space through techniques often used in angle-resolved photoemission
spectroscopy (ARPES).  Analyzing over 1200 simulations we consider
variations of the microscopic coupling strength, phonon energy and
dimensionality for two models: a momentum-independent Holstein
model, and momentum-dependent coupling to a breathing mode phonon.
In this limit we find that any `effective coupling',
$\lambda_{\text{eff}}$,
inferred from the quasiparticle renormalizations differs from the
microscopic dimensionless coupling characterizing these
Hamiltonians, $\lambda$,
and could drastically either over- or under-estimate
it depending on the particular parameters and model.
In contrast, we show that perturbation theory retains good predictive
power for low coupling and small momenta,
and that the momentum-dependence of the self-energy can
 be revealed via the relationship between velocity
 renormalization and quasiparticle strength. 
Additionally we find that (although not strictly valid) it is often
possible to infer the self-energy and bare electronic structure
through a self-consistent Kramers-Kronig bare-band fitting;  and also
that through lineshape alone, when Lorentzian, it is possible to
reliably extract the shape of the imaginary part of a
momentum-dependent self-energy without reference to the bare-band.
\end{abstract}

\pacs{\vspace{-0.4cm}71.38.-k, 79.60.-i, 74.25.Jb}

\maketitle

\section{introduction}

The many-body problem allows relatively simple interactions to
transform into a wide range of exciting yet often complicated
phenomena.  The quasiparticle picture simplifies these complications
by grouping fundamental particles and excitations together into
quasiparticles, which themselves behave in a more understandable
manner.  In this picture the real part of the self-energy represents
the energy difference from the bare particle energy, and the imaginary part
the inverse lifetime of the combined excitation.  Angle-resolved
photoemission spectroscopy (ARPES) is a well established tool for the
investigation of such electronic systems as it provides access to the
electron-removal part of the momentum-resolved spectral function
$A({\bf k},\omega)$ \cite{Damascelli:reviewPS}, which is generally
written in the form:
\begin{eqnarray}
A({\bf k},\omega)=-\frac{1}{\pi}\frac{\Sigma^{\prime\prime}({\bf
    k},\omega)}{[\omega-\varepsilon^b_{\bf 
k}-\Sigma^{\prime}({\bf k},\omega)]^2+[\Sigma^{\prime\prime}({\bf k},\omega)]^2}. \label{eqn:sf}
\end{eqnarray}
\noindent

The analysis of this extremely rich data source can be both difficult
and rewarding as it depends on both the interaction self-energy
$\Sigma({\bf k},\omega)\!=\!\Sigma^{\prime}({\bf
  k},\omega)\!+\!i\Sigma^{\prime\prime}({\bf k},\omega)$, as well
as the
single-particle electronic dispersion $\varepsilon^b_{\bf k}$ (the
so-called `bare-band').  A variety of approaches to analyzing this
spectral function have been utilized, and often focus on analysis of
either quasiparticle dispersions and their path through $({\bf
  k},\omega)$ space, or lineshape and its implications for the
structure of the self-energy.  Both methods generally cut the spectral
function into curves constant in either momentum (generating a series
of energy distribution curves [EDCs]) or energy (for a series of
momentum distribution curves [MDCs]).  
In this work, using simulations which have no experimental
  limitations, we will perform
quasiparticle analysis on EDCs (which allows the identification of a
quasiparticle peak in each ${\bf k}$ slice thereby forming a
quasiparticle dispersion, $\varepsilon^q_{\bf k}$) and self-energy
analysis on MDCs (as self-energies often show stronger energy
dependence, allowing the approximation of a constant value over a
slice of constant energy).

In quasiparticle analysis one can estimate properties such as the dispersion's velocity
$v^q_{\bf k}\!=\!\partial\varepsilon^q_{\bf k}/\partial {\bf k}$, 
effective mass $m^q_{\bf k}$, where
$1/m^q_{\bf k}\!=\!\partial^2\varepsilon^q_{\bf k}/\partial {\bf k}^2$,
and quasiparticle strength $Z^q_{\bf k}$, where $Z^q_{\bf k}\!=\!\int^q\! A({\bf
  k},\omega)d\omega$ is the integral over
the coherent part of the spectral function (this is the quasiparticle
weight only, which in a somewhat loose terminology is often referred
to as {\it quasiparticle coherence} \cite{Damascelli:reviewPS}). 
If the bare-band dispersion $\varepsilon^b_{\bf k}$ is known,
the renormalization of these properties can also be calculated. 
This concept has been used to generate an `effective coupling'
(which we will denote $\lambda_{\text{eff}}$, but which is often
denoted simply $\lambda$ in ARPES literature)
in the analysis of many complex systems, often through the
so-called `mass enhancement factor' 
$m^b_{\bf k}/m^q_{\bf k}=v^q_{\bf k}/v^b_{\bf k}=Z^q_{\bf k}=1/(1\!+\!\lambda_{\text{eff}})$.  This factor
has become a de facto standard in ARPES analysis
\cite{Damascelli:reviewPS,
  vallamoly,ZX:lambda,ingle:Sr214,Kulic20001} since,
in the Migdal/Eliashberg limit after few
approximations, it is equivalent 
to the true dimensionless
microscopic coupling found in
the Hamiltonian (denoted $\lambda$ here and in
theoretical literature)
and is expected to manifest itself in a variety of
different measurements \cite{Mahan:1981,Grimvall}.  However, the large
values sometimes measured for these renormalizations and effective couplings
 via ARPES (see, for example,
Refs.\,\onlinecite{
Nature.412.510, 
PhysRevLett.86.1070, 
Fink:lambda, 
Nature.438.474}), 
should
raise the question of this scheme's universal utility
\cite{
Kordyuk:2005, 
Giustino:2008, Nature.455.E6, PhysRevLett.100.137001, 
PhysRevB.82.064513}, 
and generally whether the
limits implied by such analysis do
apply to all systems being measured
\cite{PhysRevB.68.064408,Wojciechowski1999498,
PhysRevLett.75.1158,KKBF_short}.

Another common goal of spectral function analysis is to extract the
self-energy.  In most circumstances, under the assumption of ${\bf
  k}$-independence of the self-energy, MDCs cuts through
Eq.\,\ref{eqn:sf} reduce to a simple Lorentzian form, thus allowing a
measurement of $\Sigma^{\prime}(\omega)$ and
$\Sigma^{\prime\prime}(\omega)$ through ARPES \cite{valla, ZX:lambda,
  Damascelli:reviewRMP, campuzano, Kaminski:2005,
  Kordyuk:2005,kulic}. However not only do these methods hinge on some
assumptions and/or approximations for the bare-band $\varepsilon^b_{\bf
  k}$, but more fundamentally the problem of how momentum-dependence
in $\Sigma^{\prime}({\bf k},\omega)$ and $\Sigma^{\prime\prime}({\bf
  k},\omega)$ affects this analysis is unaddressed - even though it is
known that a Lorentzian line shape does not guarantee a
momentum-independent self-energy \cite{Randeria:DidKDep}.

Here we present a methodological study of established methods and
present some new variations using one of the most studied interactions
- that of electrons and phonons.  We generate
self-energies and spectral functions where the inclusion of
momentum-dependence and all energy scales are controlled
using the least complicated
electron-phonon interaction models possible.
However, these models
lie outside the limits of Migdal's theorem \cite{Migdal} where the
Eliashberg textbook definition of $(1+\lambda)$ renormalization is
expected to be valid.  Before we delve
into our findings for quasiparticle and  self-energy
analysis, we will first  illustrate some aspects of our chosen models and how
they are simulated.

\section{the models}

\begin{figure}[t!]
\includegraphics[width=1.0\linewidth]{\figdir/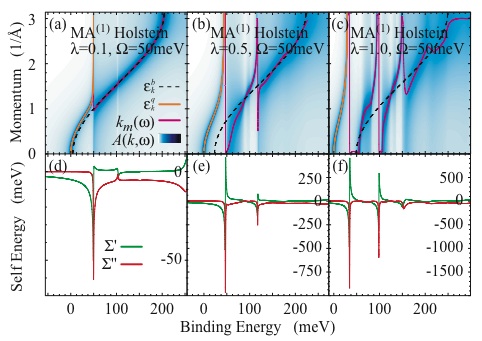}
\vspace{-0.45cm}\caption{(Color online). The spectral function (a-c) and
  self-energies (d-f) for the 1D momentum-independent
Holstein polaron model calculated with MA$^{(1)}$ for $\Omega\!=\!50$\,meV and
different microscopic couplings,
 with the bare-band $\varepsilon^b_{k}$, quasiparticle band $\varepsilon^{q}_{k}$,
 and $k_m(\omega)$ path shown.}\label{fig:lambdas_ma1}
\end{figure}

We use single electron addition to an empty
band to simulate photoemission from a completely filled system, at 0
Kelvin.  Note that this case can be exactly mapped onto
  that of a single particle added to an empty band through
  particle-hole symmetry, which essentially amounts to replacing
  $\omega \rightarrow -\omega$.  
This is an ideal test case as it provides the simplest
possible description of electron-phonon coupling and is
uncomplicated by further interactions such as strong correlations
between electrons (as in, for example, Ref.\,\onlinecite{werner}), or even a
Fermi sea which would add yet another energy scale to the problem 
(as in, for example, 
Refs.\,\onlinecite{PhysRevB.72.035125,PhysRevB.68.224511}).  The
chemical potential in our treatment is then the top of the first electron
removal state, labelled as 0 binding energy on all plots.
For momentum-independent study we will use spectral functions and
self-energies generated with the momentum-average approximation
MA$^{(1)}$ \cite{M:2006,M:2007}.
Since MA$^{(1)}$ has been shown to be 
accurate everywhere in parameter space \cite{M:2007}, it will enable
us to study $A({\bf k},\omega)$ and $\Sigma(\omega)$ over a broad
range of electron-phonon coupling and phonon energies.
For momentum-dependent study we use an extension of this
approximation with variational considerations, denoted MA$^{(v,n)}$
\cite{Goodvin2008}.  Although generally accurate everywhere in
parameter space, for reasons specific to this approximation details
studied through EDC quasiparticle analysis are best realized through
MA$^{(v,1)}$, and MDC based self-energy analysis is best realized
through MA$^{(v,0)}$ \footnote{With more terms kept exactly, 
  MA$^{(v,1)}$ should show overall improvement over MA$^{(v,0)}$,
  however for reasons which are not understood it only does so in the
  quasiparticle regime - conversely the continuum of $A(k,\omega)$
  below the quasiparticle band is worsened; toward the Brillouin zone
  edge it is pushed further down in energy than exact diagonalization
  results indicate it should be \cite{Goodvin2008,Bayo}.  As the MDC
  based self-energy analysis uses $A(k,\omega)$ in both the continuum
  and quasiparticle regime for it we use the lower order
  MA$^{(v,0)}$.}.  In all cases the spectral function remains
entirely self-consistent with the associated self-energy.

Our test case for a momentum-independent self-energy
 is the simplest possible
in momentum space, namely the Holstein polaron
\cite{Holstein:original}: momentum-independent coupling between a
single dispersionless phonon mode and 
tight-binding electrons.
In reality however, even for the Holstein model, the self-energy is
weakly dependent on momentum, which can be seen at the MA$^{(2)}$
level of approximation \cite{M:2007}.  We overcome this by choosing
the momentum-independent self-energy from the
MA$^{(1)}$ level in order to see how well these methods work for
a truly momentum-independent self energy.
For strongly momentum-dependent self-energy study
we will model coupling to a single optical mode where the phonons
live on half-integer lattice sites in between the electron sites
and modify the on-site energy of their neighbours.  In 2D this describes
lattice vibrations in a CuO$_2$-like plane, where the motion
of the O ions is the most important vibrational degree of freedom;
this has been the topic of many ARPES studies
\cite{Damascelli:reviewRMP,valla,Fink:lambda,campuzano}.  Throughout
the paper we will refer to this as the breathing-mode model.

\begin{figure*}[t!]
\includegraphics[width=1.0\linewidth]{\figdir/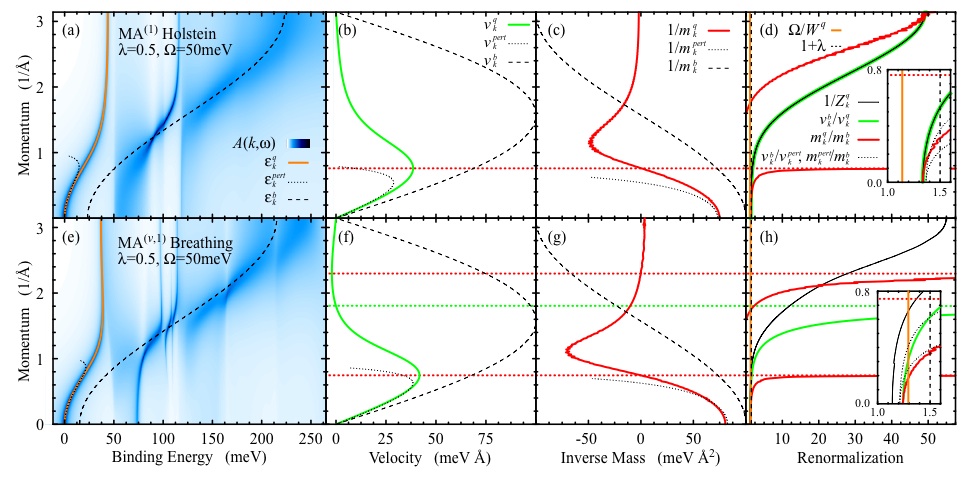}
\vspace{-0.45cm}\caption{(Color online).
(a) $A(k,\omega)$ calculated for the momentum-independent Holstein
  self-energy in 1D with MA$^{(1)}$ for $\Omega\!=\!50$\,meV and $\lambda
  \!=\!0.5$; the quasiparticle dispersion $\varepsilon^{q}_{k}$,
  perturbation theory (about $k=0$) prediction
  $\varepsilon^{pert}_{k}$, and bare-band $\varepsilon^b_{k}$ are also
  shown.  (b) Quasiparticle ($v_{k}^{q}$), perturbation theory
  ($v_{k}^{pert}$), and bare-band ($v_{k}^{b}$) velocities, as well as
  (c) corresponding inverse masses, $1/m_{k}^{q}$, $1/m_{k}^{pert}$,
  and $1/m_{k}^{b}$ according to the definitions
  $v_k\!=\!\partial\varepsilon_{k}/\partial {k}$ and
  $1/m_{k}\!=\!\partial^2\varepsilon_{k}/\partial {k}^2$.  (d)
  Momentum-dependent quasiparticle renormalization as obtained from
  $v_{k}^{b}/v_{k}^{q}$, $m_{k}^{q}/m_{k}^{b}$, as well as the inverse
  quasiparticle strength $1/Z^q_k$, where $Z^q_k\!=\!\int^q\!
  A(k,\omega)d\omega$ is the quasiparticle-only integrated spectral
  weight; in the inset, these quantities are compared near $k\!=\!0$
  to the renormalization factors $\Omega/W^q$ and $(1\!+\!\lambda)$,
  obtained from quasiparticle bandwidth $W^q$ (defined as the energy
  difference between top and bottom of the quasiparticle band) and
  dimensionless coupling $\lambda \!=\! g^2/ 2t \, \Omega $ as well as
  the perturbation theory prediction for mass and velocity
  renormalizations (shown with the same line style, but which can be
  distinguished by their proximity to the quasiparticle curves).
  In subsequent figures
 (\ref{fig:Renormalization}, \ref{fig:renorm_vary_omega},
 \ref{fig:3d_qp})
the quantity plotted is the effective coupling, $\lambda_{\text{eff}}$, which would be implied
by these renormalizations in the Migdal limit, 
which simply amounts to subtracting 1 from the renormalization.
  (e-h) demonstrate similar traces for a momentum-dependent
  self-energy
from coupling
  to a single breathing-mode in 1D, here $\lambda \!=\!
  \av{|g|^2}/2t \, \Omega$ is the average of the coupling
  across the Brillouin Zone.
The noise is due to the finite
  simulation
 grid and subsequent lineshape fitting; slight variations in peak position
are enhanced by taking the
derivative numerically and therefore most visible in
  $m_{k}^{q}$.}
\label{fig:qp_demo}
\end{figure*}

We may write both these models in the following form in momentum-space:
\begin{eqnarray}
\mathcal{H}\!=\!\sum_{\bf{k}} \varepsilon^b_{\bf{k}} c_{\bf{k}}^\dagger c_{\bf{k}}\!+\!\Omega\!\sum_{\bf{Q}} b_{\bf{Q}}^\dagger b_{\bf{Q}}\!+\!
\sum_{\bf{k},\bf{Q}} \frac{g_{\bf{Q}}}{\sqrt{N}}\!c_{\bf{k}-\bf{Q}}^\dagger c_{\bf{k}} (b_{\bf{Q}}^\dagger \!+\! b_{-\bf{Q}}). \nonumber \\
\label{eq:ham}
\end{eqnarray}
\noindent
The terms describe, in order, an electron with dispersion
$\varepsilon^b_{\bf k}=-2t \sum_{i=1}^D \text{cos}(k_i a)$ in $D$
dimensions, an optical phonon with energy $\Omega$ and momentum ${\bf
  Q}$, and the on-site momentum-dependent 
electron-phonon coupling $g_{\bf{Q}}$ [for $N$
sites with periodic boundary conditions; $c_{\bf k}^\dagger$
($c_{\bf k}$) and $b_{\bf k}^\dagger$ ($b_{\bf k}$) are the usual
electron and phonon creation (annihilation) operators].  For the
Holstein case $g_{\bf Q}\!\equiv\!g$ is a constant,
 leading to a dimensionless
coupling $\lambda \!\equiv\! g^2/ 2Dt\, \Omega $, the ratio
between lattice deformation energy $-g^2/\Omega$ and free-electron
ground state energy $-2Dt$.  For the breathing-mode
$g_{\bf Q}\!\equiv-i\sqrt{2}g\sum_{i=1}^D\text{sin}\!(Q_i a/2)$,
which has an
average value of $\langle|g_{\bf Q}|^2\rangle=g^2$ across the Brillouin
zone, allowing us to keep the same dimensionless
coupling \footnote{In most implementations, the coupling is found
  via a scattering integral around the Fermi surface
  \cite{Mahan:1981}, which for the breathing-mode model would be zero
  at all coupling strengths.  We take the Brillouin zone as a sensible
  alternative in this case (the choice is irrelevant for the
  momentum-independent Holstein coupling).}.  For this
paper we set $a=\hbar=1$ and $t=50\text{\,meV}$, such that the 1D
bandwidth is $200\,\text{meV}$ and the Brillouin zone is
$2\pi\text{\AA}^{-1}$ wide.  Also note that an additional constant
$1\,\text{meV}$ FWHM Lorentzian broadening is used, similar to an
impurity scattering, to allow the numerical resolution of the sharpest
features in $A(k,\omega)$.

The spectral function calculated with MA$^{(1)}$ for the Holstein
problem in 1D with $\Omega\!=\!50$\,meV and $\lambda \!=\!0.1, 0.5,
1.0$ is presented as a false color plot in
Fig.\,\ref{fig:lambdas_ma1}(a,b,c), along with the path of peak maxima
measured through MDCs [$k_m(\omega)$] and EDCs ($\varepsilon^{q}_{k}$)
compared with the bare-band dispersion $\varepsilon^b_{k}$.  In the
bottom panels (d,e,f) of Fig.\,\ref{fig:lambdas_ma1} we present the
real, $\Sigma^{\prime}(\omega)$, and imaginary,
$\Sigma^{\prime\prime}(\omega)$, parts of the self-energy for this
momentum-independent model.  Here each $\varepsilon^q_{k}$ is a true (and
the lowest) pole of the Green's function (it has zero width, hence an
infinite lifetime), and is only resolved numerically owing to the
impurity scattering inserted in the energy direction.  One can see
from Eq.\,\ref{eqn:sf} that the peak width should go roughly like
$\Sigma^{\prime\prime}(k,\omega)$, and it is reassuring to see in
Fig.\,\ref{fig:lambdas_ma1}(d,e,f) that the imaginary part of the
self-energy is indeed zero near $\varepsilon^q_{k}$.
The pole structure $\varepsilon^q_{k}$ is also distinct from
that of $k_m$, the path of peak
maxima observed during MDC analysis; not only are they
fundamentally different (as one is a function of $\omega$ and the
other of $k$), but the path of peak maxima observed when cutting
$A(k,\omega)$ in energy vs. momentum will not necessarily overlap, as
has already been noted \cite{ingle:Sr214,PhysRevB.67.144503}.

\begin{figure}[]
\includegraphics[width=1.0\linewidth]{\figdir/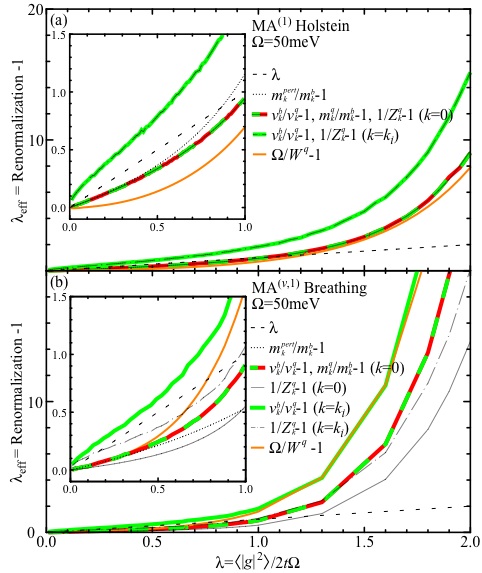}
\vspace{-0.45cm}\caption{(Color online).
Effective coupling, $\lambda_{\text{eff}}$ (as would be
interpreted in the Migdal limit from
the renormalization parameters defined as in Fig.\,\ref{fig:qp_demo}),
plotted vs. the true dimensionless coupling $\lambda \!=\! \av{|g|^2}/ 2t \, \Omega$; 
where $g$ is a constant for the
Holstein model (a), and $g_Q\!\equiv-i\sqrt{2}g\text{sin}\!(Qa/2)$ for
the breathing-mode model (b).  Also shown, in the inset only,
are the predictions for observed effective coupling
found via the mass renormalizations in perturbation theory
(Eqs.\,\ref{eq:pert_holstein_band} and \ref{eq:pert_breathing_band})
for the low-coupling regime
at $k=0$.
Note that the noise in $v$ and
$1/Z$ at $k\!=\!k_i$ originates from the numerical determination of
the inflection point
$k_i$.}\label{fig:Renormalization}
\end{figure}

For small couplings [Fig.\,\ref{fig:lambdas_ma1} panel (a)] most of
the spectral weight remains along $\varepsilon^b_{k}$, with only a
small feature formed at energy $\Omega$ below the top of the band.
With experimental resolution such a feature might appear only as a
"kink" in a quasiparticle dispersion, however from looking at the
self-energy [panel (d)] one can see that a distinction between
$\varepsilon^b_{k}$ and $k_m(\omega)$ should be made at this feature.
The lowest pole, where $\Sigma^{\prime\prime}(\omega)\approx 0$ and
which we will identify as the quasiparticle, only exists between the
top of the band and $\Omega$.  This pole forms a narrow dispersion,
$\varepsilon^q_{k}$, of bandwidth approximately $\Omega$, although for
$k$ near the zone edge the electron spectral weight is very weak due
to it having significant phonon character.  The $k_m(\omega)$ path of MDC
peak maxima, however, does not follow this quasiparticle dispersion
but instead carries on close to the original bare-band
$\varepsilon^b_{k}$ into what we will identify as the continuum due to
its broader structure and finite $\Sigma^{\prime\prime}$.

As the coupling is increased [Fig.\,\ref{fig:lambdas_ma1} panels
(b,c)],
 this distinction becomes
increasingly more evident; the quasiparticle band gains spectral
weight toward the zone boundary and becomes more well defined.  Also
its bandwidth narrows, becoming less than $\Omega$ as the
quasiparticle mass increases and the quasiparticle velocity decreases.
At the same time the spectral weight in the continuum becomes more
spread out at deep energies, and new quasiparticle-like features begin
to appear at the top of the continuum.  At very large coupling (not
shown) these additional features and the quasiparticle will eventually
form a ladder of states with flat dispersions, although this coupling
regime is well beyond the scope of this paper.

\section{Quasiparticle analysis}\label{sec:EDCs}

\begin{figure*}[t!]
\includegraphics[width=1.0\linewidth]{\figdir/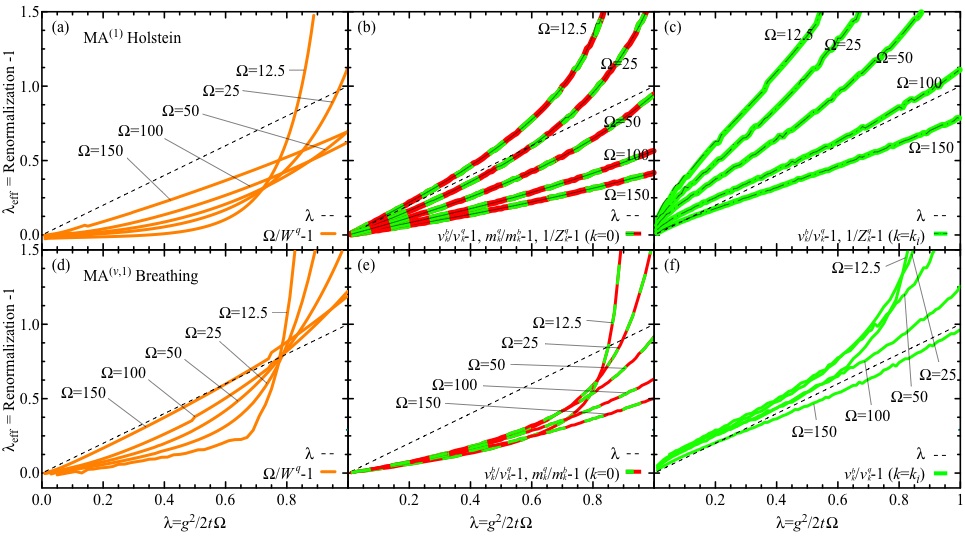}
\vspace{-0.45cm}\caption{(Color online).
Effective coupling, $\lambda_{\text{eff}}$, 
(as would be
interpreted in the Migdal limit from
the renormalization parameters
defined in Fig.\,\ref{fig:qp_demo})
from the
Holstein model (a-c) and breathing-mode model
(d-f) plotted vs. the true dimensionless coupling
$\lambda \!=\! g^2\!/ 2t \, \Omega$
for a range of phonon energies $\Omega$ labelled in meV.
In panels (b, c) those from multiple
renormalization parameters (which lie directly on top of each other)
are all plotted, whereas in (e, f) those from inverse quasiparticle weight
do not fall on any other curve and
are therefore omitted for clarity.
The slope at $\lambda=0$ in panels (b, e)
is the quantity
plotted in Fig.\,\ref{fig:pert_slopes}.
Note that the noise in panels (c, f) originates mostly
from the numerical determination of
the inflection point
$k_i$, while in (a, d) it stems mostly from variations fitting the
quasiparticle peak location at high momentum where it
is has less weight.
}\label{fig:renorm_vary_omega}
\end{figure*}

\begin{figure}[]
\includegraphics[width=1.0\linewidth]{\figdir/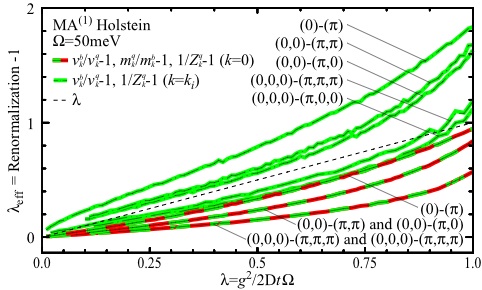}
\vspace{-0.45cm}\caption{(Color online).
Effective coupling, $\lambda_{\text{eff}}$, 
(as would be
interpreted in the Migdal limit from the mass and velocity
  renormalization parameters
  defined in Fig.\,\ref{fig:qp_demo})  from the Holstein model
 and plotted vs. the true
  dimensionless coupling $\lambda \!=\! g^2\!/ 2Dt \, \Omega$ 
 for different dimensionalities and
  high-symmetry cuts. Note that the noise in $v$ and
$1/Z$ at $k\!=\!k_i$ originates from the numerical determination of
the inflection point
$k_i$.}\label{fig:3d_qp}
\end{figure}

\begin{figure}
\includegraphics[width=1.0\linewidth]{\figdir/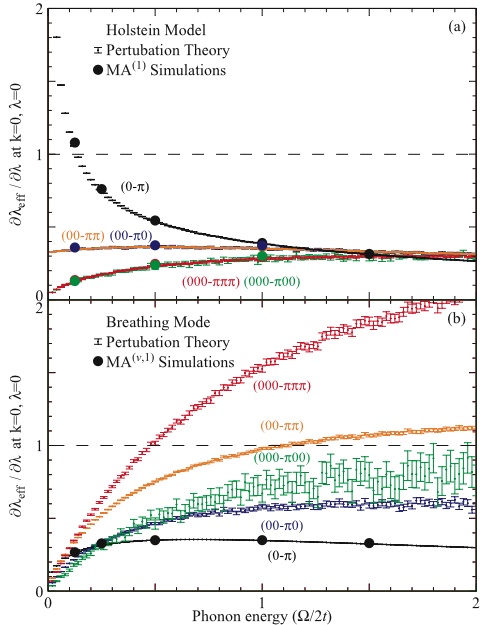}
\vspace{-0.45cm}\caption{(Color online).  Rate of
change in observed effective coupling, $\lambda_{\text{eff}}$,
per change in true dimensionless coupling, $\lambda$, 
defined as $\partial \lambda_{\text{eff}} / \partial \lambda \big|_{k=0, \lambda=0}$ 
(where $\lambda_{\text{eff}} \equiv m_{k}^{q} / m_{k}^{b} -1$
is as predicted by perturbation
theory and measured from simulated spectral functions 
and $\lambda
\!=\! \langle|g|^2\rangle/2Dt \,\Omega$) plotted
as a function of
the phonon energy for (a) Holstein model and 
(b) coupling to a breathing mode.  If the Migdal limit
holds this derivative would be a constant, 1, 
for all values of $\Omega/2t$.
For $D > 1$ cuts from the $\Gamma$ point to both the
corner and face of the Brillouin Zone were considered, as labelled.
Vertical error in the results
from simulated spectral functions is comparable to symbol size and
results from approximating
the slope at $\lambda=0$ from the finite data points
in Figs.\,\ref{fig:renorm_vary_omega} and \ref{fig:3d_qp}, as well as similar
simulation sets not shown.  Uncertainty in the perturbation
theory
results stems from the numeric Monte Carlo integration used to
determine the perturbation energies, taken from the distribution of
independent subsets of total points evaluated, and is higher for non-diagonal cuts
due to the narrower bandwidth in that direction.  The curves predicted
by Ref.\,\onlinecite{Frank_Holstein_recent} for the Holstein Model in 1 and
2 dimensions are not shown as they would be hidden by symbols,
but fall exactly onto the perturbation
theory results.}
\label{fig:pert_slopes}
\end{figure}

As can be visualized from Fig.\,\ref{fig:lambdas_ma1}, quasiparticle
renormalizations do increase as the microscopic coupling increases.
This monotonicity has led to widespread acceptance of
measuring coupling
through the quasiparticle mass,
velocity or strength renormalizations observed with ARPES,
often without
reference as to whether or not the system
should be expected to fall in
the Migdal/Eliashberg framework.
In this section we use our simple
models to demonstrate that this scheme is
not universal, and to make other
observations, by performing quasiparticle analysis as is typically done
with ARPES data (Fig.\,\ref{fig:qp_demo}) on $\sim\!1200$ generated
spectral functions.  These allow us to explore a wide range of couplings
(Fig.\,\ref{fig:Renormalization}), parameters 
(Fig.\,\ref{fig:renorm_vary_omega}), and different dimensionality 
(Fig.\,\ref{fig:3d_qp}) on models which provide both momentum-dependent and
momentum-independent self-energies.  Following a discussion of these
results we will follow the mass renormalization behavior as
$\lambda \rightarrow 0$ for ${\bf k} \sim 0$ in detail through
perturbation theory (see Fig.\,\ref{fig:pert_slopes} and
Eqs.\,\ref{eq:pert_holstein_band} and \ref{eq:pert_breathing_band}),
the predictions for which are also plotted in Figs.\,\ref{fig:qp_demo}
and \ref{fig:Renormalization} for comparison.

  In order to perform
quasiparticle analysis we generate an entire spectral function for
each combination of the parameters: model, dimensionless coupling $\lambda
\!=\! \langle|g|^2\rangle/2Dt \,\Omega$, phonon energy scale $\Omega/2t$,
dimensionality $D$, and (for 2 and 3D) the desired cut through
momentum space.  For all simulations the form of the bare band is not
changed and the hopping is set to a constant of $t=50\text{meV}$ to
give physically familiar values, a bandwidth of 200meV in the 1D case.
(To consider other bandwidths one should simply scale the bandwidth,
phonon energy, and coupling together as seen in the Hamiltonian,
Eq.\,\ref{eq:ham}).  On each of the $\sim\!1200$ generated spectral functions
the quasiparticle dispersion is found by fitting a Lorentzian peak
with linear background to each EDC within the quasiparticle regime.
The inclusion of a linear background allows the exclusion of any
spectral weight from the continuum which bleeds in (a problem
especially at low couplings and high dimensions).  We illustrate this
analysis in Fig.\,\ref{fig:qp_demo}, where we present the spectral
function $A(k,\omega)$ for a mid-range coupling $\lambda=0.5$ and
phonon energy $\Omega=50\text{meV}$ for both the Holstein and
breathing-mode models as well as dispersions found from the Lorentzian
fits $\varepsilon^q_k$, perturbation theory prediction
$\varepsilon^{pert}_k$, and the bare electronic structure
$\varepsilon^b_k$ [panels (a,e)].  Also shown are the velocities [$v_{k}^{q}$,
  $v_{k}^{pert}$ and $v_{k}^{b}$ in panels (b,f)], inverse masses
[$1/m_{k}^{q}$, $1/m_{k}^{pert}$ and $1/m_{k}^{b}$ in panels (c,g)],
the corresponding renormalization ratios
$v_{k}^{b}/v_{k}^{q}$, $m_{k}^{q}/m_{k}^{b}$ and their perturbation
theory predictions, along with the inverse quasiparticle strength
$1/Z^q_k$ and bandwidth renormalization $\Omega/W^q$ [panels (d,h),
  see caption for definitions].  In
Fig.\,\ref{fig:qp_demo} panels (a-d) present the results for the
Holstein model (with a momentum-independent self-energy),
while panels (e-h) refer to the breathing-mode coupling (with a
momentum-dependent self-energy).

Fig.\,\ref{fig:qp_demo} (d,h) show that the velocity, mass, and
spectral weight renormalizations are all functions of momentum, which
raises concerns if one would compare them to
the bandwidth renormalization $\Omega/W^q$,
or an `expected' renormalization factor of $(1+\lambda)$
which are both constant.
Although they do cross at certain values of $k$ this is merely
accidental, and none match at the top of the band - our `Fermi
surface'.  More problematic is that the mass renormalization must
necessarily contain a divergence if the inflection point of
$\varepsilon^b_k$ is different from $\varepsilon^q_k$, where
$1/m_{k}^q$ vanishes (emphasized by the horizontal dashed line).
Similarly, in the case of momentum-dependent coupling [panels (e-h)],
it can be seen that the quasiparticle dispersion $\varepsilon^q_k$ is
not even monotonic, causing another divergence when $v^q_k$ vanishes
in the middle of the dispersion (this non-monotonic dispersion is a
direct consequence of the structure of the polaronic cloud, which
causes a larger second-nearest-neighbour hopping, and is discussed at
length in Ref.\,\onlinecite{Bayo}).  Due to this momentum-dependence, any
estimation of $\lambda$ drawn from $v_{k}^{b}/v_{k}^{q}$,
$m_{k}^{q}/m_{k}^{b}$, or $1/Z^q_k$ would depend heavily on the
momentum chosen; and if either of $v_{k}^{b}/v_{k}^{q}$ or
$m_{k}^{q}/m_{k}^{b}$ were used close to their divergences, the
estimated value could be off by an unlimited amount.  Even
$\Omega/W^q$, although constant in $k$, does not match the value of
$(1+\lambda)$ for either the momentum-independent or
momentum-dependent case.
From Fig.\,\ref{fig:qp_demo} we draw the conclusion that none of the
renormalization quantities gives a good direct estimate of the
dimensionless coupling $\lambda \!=\! \langle|g^2|\rangle/ 2t \, \Omega $.
Further we find
that, with the exception of the quasiparticle strength and velocity
renormalization in the Holstein model only (which we will return to),
the renormalizations do not match even each other - even though the
models were kept as similar and simple as possible.  This indicates
that making even qualitative comparisons of
`coupling' from experiments on different materials (or even different experiments on
the same material) through these renormalization parameters may not be
meaningful.
However, modelling of the parameters in question from the original
Hamiltonian via perturbation theory might be a start,
as these results show much closer agreement near $k=0$ despite
the relatively high (for perturbation theory) coupling (we will
return to discuss perturbation theory later).

Despite their differences from each other and their momentum
dependence, however, these
renormalizations do monotonically increase as the microscopic coupling increases
(as previously observed in Fig.\,\ref{fig:lambdas_ma1})
which naturally leads one to wonder how, precisely, these quantities
scale with $\lambda$ as well as $\Omega$ in our different models, so
that one might be able to capture the trend if the material and
measured quantity is held constant - for example in an experiment
performed as a function of doping, if that doping does not cause
structural distortions.  In Fig.\,\ref{fig:Renormalization} we will
follow the `effective coupling', $\lambda_{\text{eff}}$
(which is simply the observed renormalization
minus 1), that each of these renormalization
quantities would predict using the Migdal/Eliashberg framework as a
function of $\lambda$, as well as
renormalizations found using the perturbation theory results around
$k=0$ (Eqs.\,\ref{eq:pert_holstein_band} and
\ref{eq:pert_breathing_band}),  for both momentum-independent [panel
  (a)] and momentum-dependent [panel (b)] self-energies.  In
Fig.\,\ref{fig:renorm_vary_omega} we  plot a selection of these
quantities in the same fashion, for a range of phonon energies.  For
the momentum-dependent quantities we must choose a $k$ value: we plot
$v_{0}^{b}/v_{0}^{q}$, $m_{0}^{q}/m_{0}^{b}$, and $1/Z^q_0$ at $k=0$
(our `Fermi surface'), as well as $v_{k_i}^{b}/v_{k_i}^{q}$ and
$1/Z^q_{k_i}$ at the inflection point $k\!=\!k_i$ of the quasiparticle
band $\varepsilon^{q}_{k}$, where $m_{k_i}^{q}/m_{k_i}^{b}$ diverges.

In Fig.\,\ref{fig:Renormalization} we find that the predictions from
all quantities scale
monotonically with the microscopic coupling and are concave up.  In
the low coupling regime (below about $\lambda=0.3$) the perturbation
theory results match the simulations - however nowhere does
$\lambda_{\text{eff}}$ match $\lambda$ from the Hamiltonian.
At small coupling values, using
this model, many renormalization quantities would drastically
underestimate
the true microscopic coupling, by a factor ranging from infinite
($\Omega/W$ near $\lambda=0$ where it is not renormalized in
the Holstein model) to
$\sim\!6$ ($1/Z_{k_i}$ near $\lambda=0.1$, breathing-mode).
Conversely, at larger coupling values ($\lambda\!\sim\!2$) all quantities
 would \emph{over}estimate the true microscopic
coupling, with factors ranging from $\sim\!4$ ($\Omega/W$, Holstein)
 to $\sim\!22$ ($v_{k_i}^{b}/v_{k_i}^{q}$, breathing-mode).  We
also find that, depending on the coupling or model, the relative
renormalization strength of quantities changes - for the
momentum-independent model $\Omega/W$ is renormalized the least,
whereas in the momentum-dependent model $1/Z^q_0$ shows the least
renormalization.  This indicates, yet again, that comparing different
materials via renormalizations is not feasible, nor is comparing
different renormalizations on the same material without a detailed
model.  We note again that, in the Holstein model only, quasiparticle
strength and velocity renormalization are identical for all couplings
at both $k=0$ and $k=k_i$ (as previously seen in
Fig.\,\ref{fig:qp_demo} where they are identical at all momenta).

In Fig.\,\ref{fig:renorm_vary_omega} we follow the same quantities for
a variety of phonon energies, allowing $\Omega$ to vary from
$1/16$ to $3/4$ of the bare-band width for
both models (although inverse quasiparticle weight is omitted from the
breathing-mode plots for clarity).  First we note that there are some
qualitative similarities, but just as many differences.  In all these 1D cases
the concavity increases as phonon energy decreases, so that by the
mid-coupling regime $(\lambda \approx 1)$ we recover the expected
dependence - phonons which are easier to excite (require less energy)
renormalize the band more.  However, in the low coupling regime we do
not find this dependence (later seen again in Fig.\,\ref{fig:pert_slopes} 
and in agreement with
Eqs.\,\ref{eq:pert_holstein_band}
\,and\,\ref{eq:pert_breathing_band}).
For both models
the bandwidth [panels (a, d)] shows the opposite behavior for low
coupling, with a transition near $\lambda=0.8$.  Still considering the
low-coupling regime, mass and velocity renormalizations show little
dependence on the phonon energy for the breathing-mode model, yet
strong dependence in the Holstein case.  Again we find that the
renormalizations and their corresponding `effective couplings' 
vary widely from each other, and depend on the model
and parameters chosen - sometimes in counterintuitive ways.

The final parameter to be varied is dimensionality, which we explore
briefly with Fig.\,\ref{fig:3d_qp} 
for the Holstein Hamiltonian in the low coupling regime only.
  Here we
find that for a fixed dimensionality and phonon energy where $\Omega \sim t$
the renormalizations as a function of $\lambda$ look qualitatively
 similar.  The various
renormalized quantities increase monotonically yet 
 remain distinct from the microscopic coupling as well as
each other (with the exception of quasiparticle
strength and velocity renormalization which are again the same),
with details that depend on phonon energy and dimensionality.  We
feel it is important to note, however, that at larger couplings
not explored here
other studies on the dynamics of the Holstein 
(and momentum dependent Su-Schrieffer-Heeger) models
have found more
complicated behavior in higher
dimensions; where a critical coupling value marks a drastic change
in quasiparticle properties, which is most prominent 
as $\Omega \rightarrow 0$
\cite{Frank_Holstein_recent,PhysRevB.81.165113,EurophysLett.42.523,PhysRevB.56.4484}.
However interesting,
this type of
behavior would not simplify quasiparticle
renormalization analysis on such a system and is not investigated here.

So far we have shown that, while the slope may not be 1,
the renormalization curves could all still be reasonably well approximated
as \emph{linear} in $\lambda$ in the very low coupling regime and that perturbation theory
makes an excellent prediction for them near $k=0$.
This allows us to follow
this slope more continuously through parameter space with
perturbation theory than by simulating even larger numbers of spectral functions.
It is worth noting that in the classic implementation of the
$(1+\lambda)$ scheme (see Ref.\,\onlinecite{Grimvall}), perturbation theory
is discussed but dismissed as a possible avenue due to the resulting
corrections being too large for perturbation theory to be valid.
However, in that instance, some approximations are made to ease
completion of the integrals which eliminate the possibility of the
arbitrarily small couplings we have used here.
In our case the lowest non-zero order in the
phonon-electron interaction term from Eq.\,\ref{eq:ham} is the second, and
in 1D we find that for both models it is possible to complete the integrals
without further approximation.  In the Holstein case the energy
dispersion should be modified from $\varepsilon^b_k =
-2t\text{cos}(ka)$ to:
\begin{equation}
\varepsilon^{pert}_k \approx -2t \left( \text{cos}(ka) 
+ \lambda \frac{ \frac{\Omega}{2t} }{\sqrt{\left( \text{cos}(ka) + \frac{\Omega}{2t} \right)^2 - 1}}\right),
\label{eq:pert_holstein_band}
\end{equation}
which agrees with the results calculated for the quasiparticle residue
at $k=0$ in Ref.\,\onlinecite{Frank_Holstein_pert}.  For the breathing-mode
model we find that the dispersion becomes:

\begin{eqnarray}
\varepsilon^{pert}_k & \approx & -2t \left( \text{cos}(ka) 
+ \lambda \frac{\Omega}{2t} \, \mathcal{F}\left( \textstyle{\frac{\Omega}{2t}},k \right)  \right) \nonumber \\
\mathcal{F}\left( \textstyle{\frac{\Omega}{2t}},k \right) & \equiv &
\text{cos}(ka) + \frac{\text{sin}^2(ka) - \frac{\Omega}{2t} \text{cos}(ka)}
{\sqrt{\left( \text{cos}(ka) + \frac{\Omega}{2t} \right)^2 - 1}}.
\label{eq:pert_breathing_band}
\end{eqnarray}

This demonstrates that, at the very least, we should not expect the
renormalizations to depend solely on the dimensionless coupling
$\lambda=g^2/2t\Omega$, but also on the other relevant energy scale in
the problem - the ratio of phonon energy to bandwidth.  By taking
derivatives of these dispersions we can also find the predicted mass
and velocity renormalizations.  In Fig.\,\ref{fig:qp_demo} we plot the
predicted dispersion, derivatives, and renormalizations contrasted
against the simulated spectral function and find close but not perfect
agreement for both models near $k=0$ (but failing at large momenta),
despite the relatively high coupling ($\lambda=0.5$).  As seen in
Figs.\,\ref{fig:Renormalization} and \ref{fig:pert_slopes}
near $k=0$ for vanishing $\lambda$ there 
is perfect agreement within our measurement accuracy;
perturbation theory begins to show signs of
failure near $\lambda \approx 0.25$.
In higher dimensions we did not complete the integrals exactly but
instead used the VEGAS Monte Carlo integration algorithm to
evaluate them numerically \cite{vegas1,vegas2,gsl}.  Using this routine
for all dimensionalities allowed us to validate the results by comparing
them to these known solutions for 1D for both models and the results
found in
Refs.\,\onlinecite{Frank_Holstein_pert,Frank_Holstein_recent} for the
Holstein model in 1 and 2D, where they show perfect agreement
(once corrected for a slightly different definition of $\lambda$ in
2D).
 
In Fig.\,\ref{fig:pert_slopes} we show how the renormalization with
the dimensionless coupling parameter $\lambda$ near $k=0$,
$\lambda=0$  (and hence the observed effective coupling,
$\lambda_{\text{eff}}$)
varies as a
function of the other energy scale $\Omega/2t$ for both models in
1, 2 and 3 dimensions, and
how this matches nearly perfectly against measurements of the
same quantity on the simulations. 
Interestingly, despite spectral functions which have
outwardly similar characteristics [as seen, for example, in
  Fig.\,\ref{fig:qp_demo} panels (a vs. e) or
  Fig.\,\ref{fig:kkbf_momdep} vs. Fig.\,\ref{fig:kkbf_results}], we
find a drastic difference in how the actual renormalizations vary with phonon
energy depending on the model, and that neither model would be well
approximated by a $\lambda_{\text{eff}}=\lambda$ scheme, which is
shown as the dashed line along 1.
The 1D Holstein model shows a perhaps expected dependence:
phonons which require very little energy to excite have a dramatic
effect on the electronic renormalization (blowing up as $\Omega
\rightarrow 0$); but 
as the phonon energy increases, the
 mode has progressively less effect. In the 2D case, however,
we find very limited dependence on
phonon energy with a curve that is almost flat and could 
therefore be rescaled to match if
$\lambda$ was chosen to be defined appropriately.  In 3D we find the
opposite of the 1D case whereby the renormalization
vanishes as $\Omega \rightarrow 0$.
These very different limits are directly related to the bare-electron
DOS at the band-edge, and its strong dependence on
dimensionality \cite{Frank_Holstein_recent}.
In all dimensions we find that the
renormalization is isotropic (as one might expect from an isotropic
coupling) and that it asymptotically approaches a similar value
for large phonon energies - reminiscent of a renormalization
which depends solely on $\lambda$, if only for $\Omega/2t \gg 1$.
In stark contrast, however, the more realistic breathing-mode model
shows entirely different behavior.  For all dimensionalities the
overall functional form is similar for $\Omega \rightarrow 0$,
where mass renormalizations vanish.
This low energy behavior may be expected as
for weak coupling and energies close to $k=0$ in the bare-band, 
the electron couples mostly to
 $q\approx 0$ phonons; and in this model such coupling vanishes,
  $g(q\rightarrow 0)
  \rightarrow 0$.  As the phonon energy increases, however, we
discover that the renormalization is anisotropic with
stronger renormalization along the diagonal
cut (as may be expected for an anisotropic coupling) and a coupling
which gets stronger as dimensionality increases (opposite the
Holstein case).  We also find
that the renormalizations do not asymptotically approach any fixed
value for large phonon energies, as they did for the Holstein case.

Overall we find that there is much variability in quasiparticle
analysis, to the point that one cannot make a general rule about
renormalizations in this regime.  There
are, however, two common threads.  Firstly, for both our models,
perturbation theory works in the low coupling regime: it correctly
predicts the quasiparticle band structure near $k=0$ for all
combinations of parameters tested, although it fails at higher
momenta (as seen in Fig.\,\ref{fig:qp_demo}).  The second, and
perhaps more interesting, hints at something which may be
quantitatively gained through quasiparticle analysis - without even a
more detailed model on which to attempt perturbation theory or other
tools.  In Figs.\,\ref{fig:qp_demo}, \ref{fig:Renormalization},
\ref{fig:renorm_vary_omega}, and \ref{fig:3d_qp} one observes that, in
the momentum-independent case only, $v_{k}^{b}/v_{k}^{q}$ and
$1/Z^q_k$ lie precisely on top of each other for all values, and match
$m_{k}^{q}/m_{k}^{b}$ at $k=0$.  Although the velocity and mass
renormalizations at $k=0$ are simply a consequence of derivatives
following each other near an extremum, the velocity renormalization
and quasiparticle strength have implications for the structure of the
self-energy, as was previously noted in Ref.\,\onlinecite{KKBF_short} and
is discussed in greater detail now.

By definition, the Green's function is:
\begin{equation}
G(k,\omega) = \frac{1}{\omega - \varepsilon^{b}_{k} - \Sigma(k,\omega)
  + i\eta}.
\end{equation}
In the infinite lifetime quasiparticle regime the self-energy should
be purely real, with any broadening coming solely from the small
impurity scattering, $\eta$.  We may then identify the implicitly
defined quasiparticle dispersion as $\varepsilon^{q}_{k} =
\varepsilon^{b}_{k} +
\Sigma(k,\omega)\big|_{\omega=\varepsilon^{q}_{k}}$ and, since we are
interested in an EDC, expand the self-energy about
$\omega=\varepsilon^{q}_{k}$ to first-order in energy.  Taking $-1/\pi$
times the imaginary part yields the spectral function:
\begin{equation}
A(k,\omega) \approx \frac{1}{\pi} \frac{\eta}{\eta^2 +
  (\omega-\varepsilon^{q}_{k})^2(1 -
  \frac{\partial\Sigma(k,\omega)}{\partial\omega}
  \big|_{\omega=\varepsilon^{q}_{k}} )^2}.
\end{equation}
We can see that, cut along energy in the quasiparticle regime,
the spectral function will be a Lorentzian at $\varepsilon^q_k$ with
width given by $\eta$ and with weight $Z_k=1/(1
-\frac{\partial\Sigma(k,\omega)}{\partial\omega}
\big|_{\omega=\varepsilon^{q}_{k}})$.  This relationship between
quasiparticle weight and the energy derivative of the self-energy is
often derived, and usually associated directly with the success of 
an effective
coupling scheme \cite{Grimvall, Mahan:1981}, but we do \emph{not} make that
association here.  Velocity renormalization is simply the ratio of
momentum derivatives of the bare, $v^b_k$, and quasiparticle
$v^{q}_{k} = v^{b}_{k} +
\frac{\partial\Sigma(k,\omega)}{\partial\omega}\big|_{\omega=\varepsilon^{q}_{k}}
v^{q}_{k} + \frac{\partial\Sigma(k,\varepsilon^{q}_{k})}{\partial k}$
bands, which reduces to:
\begin{equation}
\frac{v^{b}_{k}}{v^{q}_{k}} = \frac{1}{Z_k} - 
\frac{\partial\Sigma(k,\varepsilon^{q}_{k})}{\partial k}\frac{1}{v^{q}_{k}}.
\label{eq:velorenorm}
\end{equation}

We see that, for momentum-independent self-energies, the velocity
renormalization must follow the inverse spectral weight.  This means
that the renormalization quantities can be used to conclusively check
the momentum-dependence of the self-energy, in the quasiparticle
regime.  Whether or not the self-energy is momentum-dependent is of
great importance to MDC self-energy analysis, on which we focus
in the
rest of the paper.

\section{Self-energy analysis}\label{sec:MDCs}

Since quasiparticle analysis fails to reveal the true microscopic
coupling through renormalizations, we look toward other options for
spectral function analysis.  In this section we describe how it is
possible to estimate the self-energy through the analysis of MDC
lineshapes.  We will begin, for simplicity, with a description for
momentum-independent self-energy and move on to describe the
implications of momentum-dependence on the procedure.  Cases of a
momentum-independent self-energy can be verified through quasiparticle
analysis; as seen in Eq.\,\ref{eq:velorenorm} the {\it overlap} of
$v_{k}^{b}/v_{k}^{q}$ and $1/Z^q_{k}$ is only possible when the
self-energy is {\it momentum-independent} along the quasiparticle
dispersion.  Although the quasiparticle dispersion and the path of MDC
peak maxima where MDC analysis is carried out may vary, in practice
they are often very close in the quasiparticle regime.  One must
always keep in mind that although a momentum-independent self-energy
causes a Lorentzian MDC lineshape, Lorentzian lineshape alone is not sufficient
to conclude that $\Sigma(k,\omega)\!=\!\Sigma(\omega)$
\cite{Randeria:DidKDep}.

\begin{figure}[t!]
\includegraphics[width=1\linewidth]{\figdir/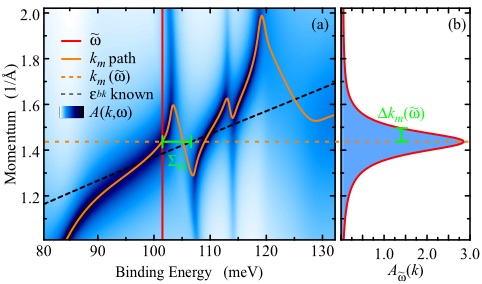}
\vspace{-0.45cm}
\caption{(Color online). Eq.\,\ref{eqn:se_lorentzian} and
  Eq.\,\ref{eqn:mdc_params} described diagrammatically for a momentum
  independent self-energy.  (a) is a false color plot with the
  bare-band ($\varepsilon^b_k$) and path of peak maxima ($k_m$ path)
  shown in addition to an example MDC cut at $\tilde\omega$ and the
  peak maximum location for that cut [$k_m(\tilde\omega)$].  (b) the
  cut through momentum of $A(k,\omega)$ at constant energy
  $\tilde\omega$, observed to be a Lorentzian with a peak maximum
  located at $k_m(\tilde\omega)$, a HWHM $\Delta k_m (\tilde\omega)$,
  and an area $A_0(\tilde\omega)$.  These lineshape properties are
  related to the self-energy at $\tilde\omega$ through the bare-band
  evaluated at $k_m$ through Eq.\,\ref{eqn:mdc_params}}
\label{fig:MDC_method}
\end{figure}

In cases where the self-energy is momentum-independent we may analyze
$A(k,\omega)$ in terms of MDCs at constant energy
$\omega=\tilde\omega$, where the self-energy may then also be
considered a constant.  Under this condition, as long as
$\varepsilon^b_{k}$ can be linearized in the vicinity of the MDC peak
maximum observed at $k=k_m$, the MDC lineshape will be Lorentzian.  By
Taylor expanding $\varepsilon^b_{k}$ about an MDC peak maximum at
$k=k_m$, i.e.
$\varepsilon^b_{k}\!=\!\varepsilon^b_{k_m}\!+\!v^b_{k_m}\!\cdot
(k-k_m)\!+\!...$, ignoring higher order terms (which must be
negligible if the curve appears Lorentzian), and noticing that
$\varepsilon^b_{k_m}\!+\!\Sigma^{\prime}_{\tilde\omega}-\tilde\omega=0$
will implicitly define the observed peak maximum, we can rewrite
Eq.\,\ref{eqn:sf} as:
\begin{equation}
A_{\tilde\omega}(k)\!\backsimeq\! \frac{A_0}{\pi} \frac{ \Delta k_m }
{ (k-k_m)^2 + (\Delta k_m)^2 }, 
\label{eqn:se_lorentzian}
\end{equation}
with:
\begin{eqnarray}
\Delta k_m = & - \Sigma^{\prime\prime}_{\tilde\omega}/v^b_{k_m} & =
\text{HWHM}, \nonumber \\
A_0 = & 1/v^b_{k_m} & = \int\!A_{\tilde\omega}(k)dk.
\end{eqnarray}
Here $\Delta k_m$ is the half-width half-maximum (HWHM) of a
Lorentzian of weight $A_0$.  For each constant energy cut,
$\omega=\tilde\omega$, the observed peak maximum is labelled $k_m$.
The self-energies are then easily found as:
\begin{eqnarray}
\Sigma^{\prime}_{\tilde\omega}&=&\tilde\omega - \varepsilon^b_{k_m},
\nonumber \\ \Sigma^{\prime\prime}_{\tilde\omega}&=&-\Delta k_m
v^b_{k_m}.
\label{eqn:mdc_params}
\end{eqnarray}
One must be careful visualizing the relationships; although a simple
picture might be that the peak, whose width scales with the imaginary
self-energy and band velocity, has been `pushed up' by the real
self-energy to its observed location at $\tilde\omega$, one must
remember that these quantities are defined implicitly and evaluated at
different locations in the $(k,\omega)$ plane: the self-energy is
evaluated at $(k_m,\tilde\omega)$ and the bare-band at
$(k_m,\varepsilon^b_{k_m})$.  Of course, these implicit definitions
are no trouble if you simply wish to {\it observe} $A(k,\omega)$ and
not {\it calculate} it based on this simple approximation.  These
relations are  illustrated graphically
in Fig.\,\ref{fig:MDC_method}. 

\begin{figure}[t!]
\includegraphics[width=1.0\linewidth]{\figdir/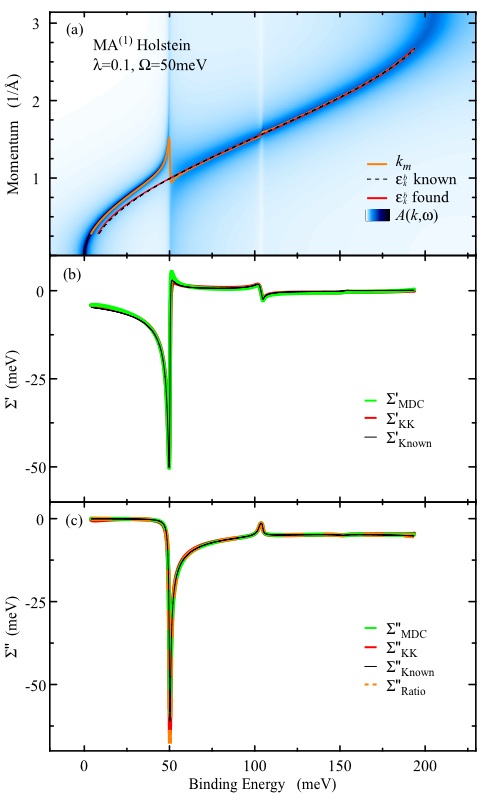}
\vspace{-0.45cm}\caption{(Color online). (a) $A(k,\omega)$ calculated
  for the momentum-independent Holstein self-energy with MA$^{(1)}$ for
  $\Omega\!=\!50$\,meV and $\lambda \!=\!0.1$ as a false color plot;
  also shown are the $k_m$ path of MDC maxima along which the analysis
  is performed, as well as the known bare-band and the third order
  polynomial approximation found through the KKBF analysis (the
  arbitrary energy offset introduced through KKBF has been shifted
  back by hand to allow comparison between the two).  (b,c) Real and
  imaginary parts of the self-energy from the model ($\Sigma_{known}$),
  the bare-band and MDC fitting routine ($\Sigma_{MDC}$) as found via
  Eq.\,\ref{eqn:mdc_params}, and the KK transform of
  $\Sigma^{\prime\prime}_{MDC}$ ($\Sigma^{\prime}_{KK}$) and
  $\Sigma^{\prime}_{MDC}$ ($\Sigma^{\prime\prime}_{KK}$) used as an
  internal check in KKBF.  In (c) the MDC ratio results,
  $\Sigma^{\prime\prime}_{ratio},$ as found via
  Eq.\,\ref{eqn:imaginary_ratio} are also shown.}
\label{fig:kkbf_results}
\end{figure}

These relationships work exactly where they are applicable: namely
when the self-energy is momentum independent, $k_m$ is far from a band
edge (where $v^b_{k_m}$ must vanish), where the peak shape is truly
Lorentzian, and when the peak width is narrow enough that a
first-order expansion of $\varepsilon^b_k$ is appropriate.  Since the
convolution of two Lorentzians is another Lorentzian
where the peak width is a simple sum of the widths of the original
functions, the inserted impurity scattering shows up directly as an
addition to the measured $\Sigma^{\prime\prime}_{\tilde\omega}$ (for
comparison purposes a constant $\eta=1$\,meV has therefore
 been subtracted
from all plots of $\Sigma^{\prime\prime}_{\tilde\omega}$). However,
these relationships still hinge on knowledge of the bare-band.  If $\varepsilon^b_k$
is unknown it is possible to fit it, to within an arbitrary
energy offset, to any functional form which provides a value and
derivative using a Kramers-Kronig bare-band fitting (KKBF) routine
(see Appendix).  Alternatively, as previously noted in
Ref.\,\onlinecite{KKBF_short} and used in Ref.\,\onlinecite{used_ratio}, the
imaginary part of the self-energy requires knowledge of only
$v^b_{k_m}$, which can be obtained directly from
$A_0\!=\!1/v_{k_m}^b$, allowing us to write it as the MDC
width/integral ratio:
\begin{equation}
\Sigma^{\prime\prime}_{ratio}\!=\!-\Delta k_m/A_{0}
\label{eqn:imaginary_ratio}
\end{equation}

This variation allows us to tackle the problem over a larger range of
$\lambda$ as it does not rely on the KKBF routine to succeed over the
entire range in order to ensure the fitness of the Kramers-Kronig
transform and fit the bare-band. Eq.\,\ref{eqn:imaginary_ratio} is
free
to work over energies
where the peak is Lorentzian (i.e. Eq.\,\ref{eqn:se_lorentzian} holds),
and to fail in others without allowing these failures to block the
procedure.  Experimentally, when using this ratio, one must be careful 
that the observed spectra are properly normalized, otherwise it will be
off by a constant multiple, but even if this is not possible the form
of the imaginary self-energy should be nevertheless recoverable.  It
is also possible, in cases of momentum-independent self-energy for which
$v_{k}^{b}/v_{k}^{q}\!=\!1/Z^q_k$ from Eq.\,\ref{eq:velorenorm}, to
find the same ratio using only quasiparticle properties as
$\Sigma^{\prime\prime}_{MDC}\!=\!-v_{k_m}^{q}\Delta k_m/Z^q_{k_m}$.

\begin{figure}[t!]
\includegraphics[width=1.0\linewidth]{\figdir/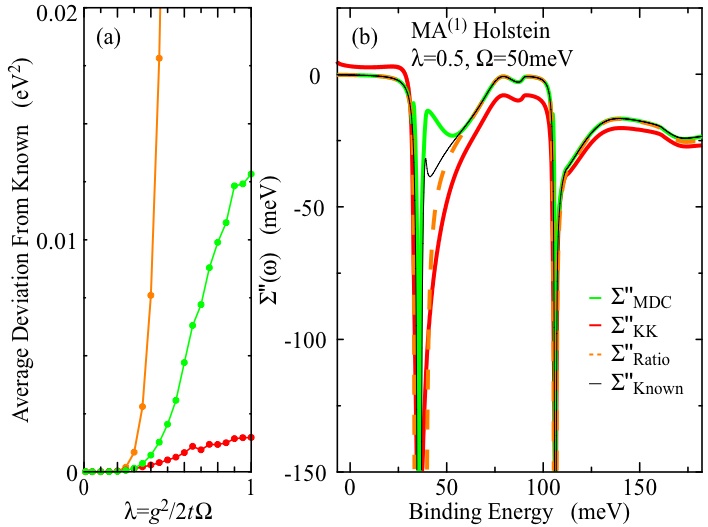}
\vspace{-0.45cm}\caption{(Color online). (a) Deviation (i.e., average
  of the squared difference at each $k_m$) between estimated and known
  self-energies vs. $\lambda$.  (b) Various estimates for the
  imaginary part of the self-energy, defined as in the caption of
  Fig.\,\ref{fig:kkbf_results}, for $A(k,\omega)$ calculated within
  MA$^{(1)}$ \,for\, $\Omega\!=\!50$\,meV and $\lambda
  \!=\!0.5$.\,}\label{fig:kkbf_failure}
\end{figure}

The results of both the KKBF and the ratio method, performed as if the
bare-band was not known on a momentum-independent self-energy in the
low-coupling regime, are presented in Fig.\,\ref{fig:kkbf_results}.
The internal self-consistency of the KKBF results is confirmed by the
good match between $\Sigma_{MDC}$ and $\Sigma_{KK}$, and the agreement
of $\Sigma_{Ratio}$ adds further confidence.  These `measured' quantities
show good agreement with their known counterparts everywhere
Eq.\,\ref{eqn:se_lorentzian} is applicable, demonstrating that these
methods work well in the low-coupling regime; however, they become
progressively less accurate as the coupling increases.  In
Fig.\,\ref{fig:kkbf_failure}(a) we show the progressive failure of the
method applied for couplings where $\lambda$ ranges from 0-1, which
demonstrates a rapid degeneration of the accuracy of the method
outside of the low-coupling regime.  Note, however, that the two
methods fail in different ways.  $\Sigma^{\prime\prime}_{MDC}$ tends
to fail more globally, whereas $\Sigma^{\prime\prime}_{Ratio}$ often
continues to work almost exactly in some energy regions while failing
catastrophically in others (these regions cause its average deviation,
shown in Fig.\,\ref{fig:kkbf_failure}, to indicate perhaps a higher
degree of failure than deserved).  In Fig.\,\ref{fig:kkbf_failure}(b)
we demonstrate these differences by showing the results of the methods
 applied blindly at $\lambda=0.5$, even though lineshapes show that
there are areas where Eq.\,\ref{eqn:se_lorentzian} does not apply.
One can see how the internal KKBF check has begun to fail as
$\Sigma^{\prime\prime}_{MDC}$ and $\Sigma^{\prime\prime}_{KK}$ do not
match; there are structural differences and
$\Sigma^{\prime\prime}_{KK}$ has picked up different offsets in the
different flatter parts of the spectrum, making setting its overall
offset difficult.  As the disagreement between
$\Sigma^{\prime\prime}_{MDC}$ and $\Sigma^{\prime\prime}_{KK}$
increases with coupling it will eventually cause the KKBF routine to
fail to capture the bare electronic structure.  None of the methods
reproduce $\Sigma^{\prime\prime}_{Known}$ near the sharp one-phonon
structure; note that $\Sigma^{\prime\prime}_{MDC}$ and
$\Sigma^{\prime\prime}_{Ratio}$ overestimate and underestimate it,
respectively.  Our experience with this model leads us to believe this
to be typical behavior: when each method fails they do not tend to
fail in identical ways, so that in regions where they do match one can
still have confidence that the methods are working.

These methods hinge on the momentum-independence of the self-energy in
two ways.  For fitting lineshape, a
momentum-independent self-energy guarantees a Lorentzian lineshape but
the inverse is not true - it is still possible to have a
momentum-dependent self-energy which generates a Lorentzian.
Additionally, fitting the bare-band relies on the Kramers-Kronig
transforms in energy, which are only valid for a fixed momentum.
In cases where the momentum-dependence is not too strong locally near
$k_m$, however, we have found that it is still possible to gain
insight using similar approaches, although even more care must be taken in
the interpretation of the results.  If, despite momentum-dependence,
the MDC appears Lorentzian in shape, certain higher order terms must be
small when expanding each of $\varepsilon^b_k$,
$\Sigma^{\prime}(\omega,k)$ and $\Sigma^{\prime\prime}(\omega,k)$
about $k_m$.  Under this condition we may drop terms of order
$(k-k_m)^3$ and higher from the denominator as well as $(k-k_m)$ and
higher from the numerator, which also implies we may drop $\partial
\Sigma^{\prime\prime} / \partial k$ and $\partial^2
\Sigma^{\prime\prime} / \partial k^2$ everywhere.  We end up with an
expression identical to Eq.\,\ref{eqn:se_lorentzian}, only with new
interpretations for the HWHM as well as the spectral weight of the
peak:

\begin{eqnarray}
\Delta k_m &=& - \frac{\Sigma^{\prime\prime} (\tilde\omega,k_m) }
                      {v^b_{k_m} +
                        \partial \Sigma^{\prime}(\tilde\omega,k_m)
                        / \partial k}, \nonumber \\
A_0 = \int\!A_{\tilde\omega}(k)dk & = & 
\frac{1}{v^b_{k_m} + \partial \Sigma^{\prime}(\tilde\omega,k_m) / \partial k}.
\label{eqn:momdep_lorentzian}
\end{eqnarray}

\begin{figure*}
\includegraphics[width=1\linewidth]{\figdir/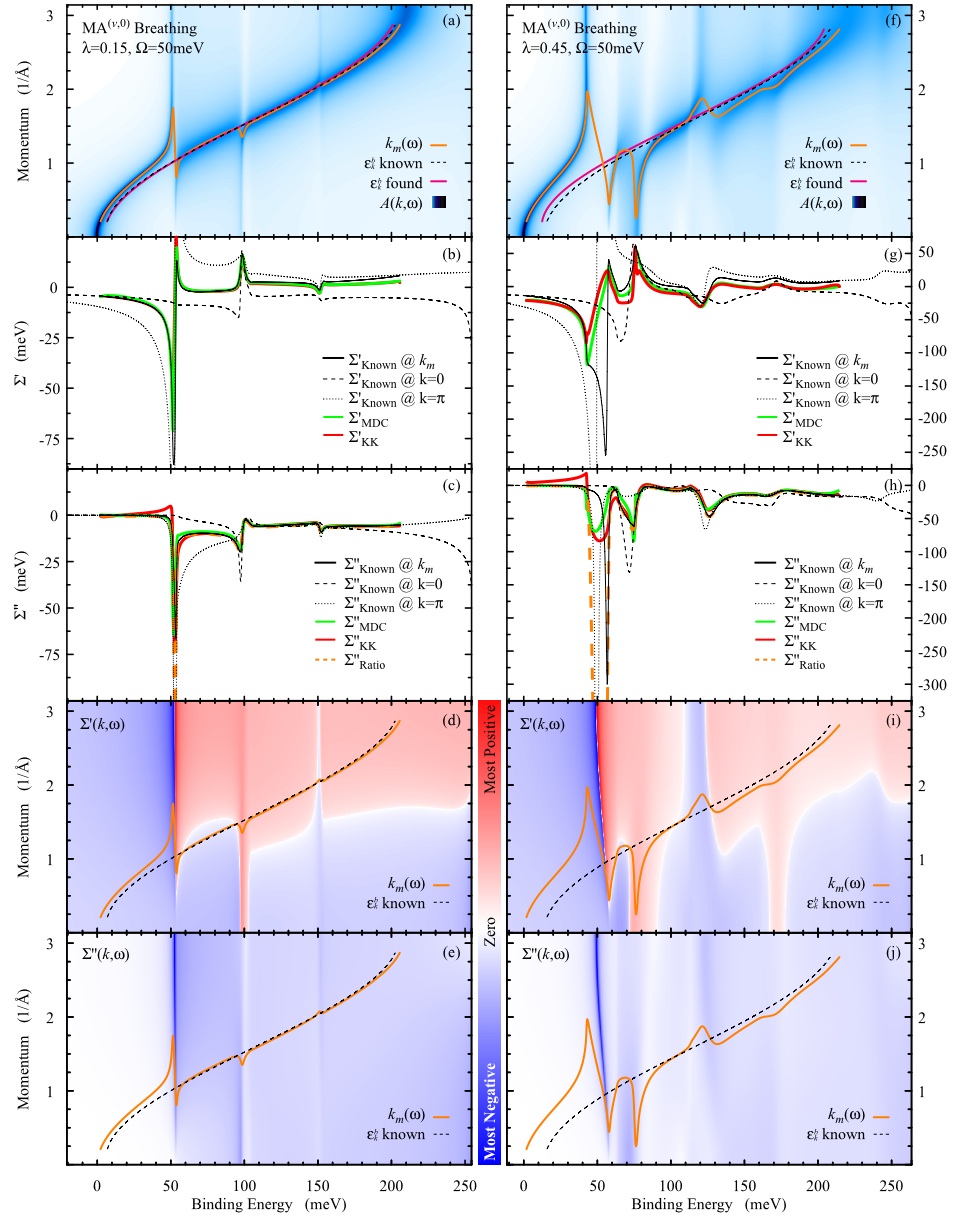}
\vspace{-0.45cm}\caption{(Color online).
(a-c,f-h) $A(k,\omega)$ and self-energies as defined in
  Fig.\,\ref{fig:kkbf_results} for momentum 
dependent coupling to a single breathing mode.
In this instance, as the self-energy is momentum-dependent, the known self-energies are
plotted along the path of peak maxima ($\Sigma_{\text{Known}} @ k_m$) to show good agreement
with the derived values, as well as along both edges of the Brillouin Zone
for comparison ($\Sigma_{\text{Known}} @ k=0$ and $\Sigma_{\text{Known}} @ k=\pi$).
Finally (d,e,i,j) show the full extent of the energy and momentum-dependence of the 
real and imaginary self-energies as a false color plot, with the $k_m$ path superimposed.}
\label{fig:kkbf_momdep}
\end{figure*}

Now that the self-energies are momentum-dependent it becomes more
important to remember that the self-energy extracted using this method
will follow the $k_m$ path through $(\omega,k)$ space; this path is
demonstrated as a false color plot in
Fig.\,\ref{fig:kkbf_momdep}(d,e,i,j).  From
Eq.\,\ref{eqn:momdep_lorentzian} we find that, in this
momentum-dependent case, the ratio check
Eq.\,\ref{eqn:imaginary_ratio} proves to be invaluable as it removes
the possible inaccuracies when strong momentum-dependence near $k_m$
in $\Sigma^{\prime}$ might provide a Lorentzian lineshape with a
misleading width viewed from a momentum-independent perspective.  In
Fig.\,\ref{fig:kkbf_momdep}(a-c,f-h) we present the results of both
KKBF and the ratio method for a momentum-dependent coupling.  From a
comparison between the measured and known
self-energies on paths through $k$ space along the zone boundaries and
along $k_m$, we find agreement only along the $k_m$ path, as expected.
Further, especially in Fig.\,\ref{fig:kkbf_momdep}h near the 3rd phonon structure close
to $\sim\!125\text{meV}$, one can see how it is possible for
$\Sigma^{\prime\prime}_{MDC}$ and $\Sigma^{\prime\prime}_{KK}$ to
agree with each other and yet not correctly predict
$\Sigma^{\prime\prime}_{Known}$, despite the peak shape being
reasonably Lorentzian, due to sufficient local momentum-dependence in
the real self-energy (Fig.\,\ref{fig:kkbf_momdep}i).  In this location we note, however,
that $\Sigma^{\prime\prime}_{Ratio}$ still correctly predicts
$\Sigma^{\prime\prime}_{Known}$ as it is not affected by this local
momentum-dependence.  Overall we find that, in a similar fashion to
the momentum-independent case, there is generally good agreement
between the found self-energies and the self-energy along the $k_m$
path in the low-coupling regime and the methods progressively fail as
we move into the mid-coupling regime.  The available modes of failure
are increased: there are more locations where the lineshape is not
Lorentzian due to strong local momentum-dependence of the self-energy;
places where it is still Lorentzian but with a misleading width; and
the Kramers-Kronig relations are not valid along an arbitrary path
through $(\omega,k)$ space, which disrupts the fitness of the KKBF
routine.  While it is not surprising that the KKBF routine eventually
fails for large couplings in the momentum dependent case it is interesting
that it works at all, as the Kramers-Kronig relations in energy are only
formally valid for a fixed momentum but the measured self-energies
follow the $k_m$ path at all couplings.  Despite this, as can be seen
in Fig.\,\ref{fig:kkbf_momdep} (b, c), the Kramers-Kronig relations
appear to work relatively well along the $k_m$ path in the low
coupling case. Nevertheless we find that, in this model, failures occur at
qualitatively similar couplings when momentum-dependence is added.

\section{Conclusions}

The spectral function is an extremely rich data source.  We have shown
that, despite its allure, away from the Migdal limit 
it is not always possible to extract the true
microscopic coupling through quasiparticle renormalizations of ARPES
data with an effective coupling scheme - and certainly
not for cases close to a dispersion maximum.  In this limit 
$\lambda_{\text{eff}} \ne \lambda$. However, through careful
modeling and the analysis of specific features it may be possible to
learn much more.  If one can correctly guess the model it may be
possible to predict a given renormalization, or even show experimental
support for a given model via relationships between renormalization
parameters. Through MDC self-energy analysis, we have shown that the
self-energy can be extracted along paths through $(k,\omega)$ space in
the low coupling regime - and it is possible still to gain some
insight at higher couplings.  We have also shown that it is possible to
infer the momentum-dependence of the self-energy through
comparison of renormalization properties.  Methods like these,
together with detailed models and powerful simulations, will hold the
key to more thorough and quantitative analysis of the rich data
supplied through ARPES.

\section{Appendix on KKBF}

The method outlined here varies slightly from techniques previously described in the literature, which
generally reduce the functional form for $\varepsilon^b_{k}$ substantially in order to 
facilitate an exact solution for $A(k,\omega)$ as they often
deal with data very close to the Fermi energy over a small range
 \cite{Damascelli:reviewRMP, campuzano,Kordyuk:2005,Kaminski:2005}.
In our analysis we have instead expanded
everything about $k_m$, essentially using a new linear approximation for $\varepsilon^b_{k}$ on 
each MDC slice.  Although our method has the disadvantage that it does not work as well near the zone boundary
where the band velocity goes to zero (other methods which make a second order approximation can successfully
predict and fit the non-Lorentzian shape in this region and may continue to work in this regime),
ours has the distinct advantage that it works over a much larger energy range and allows fitting based on an
infinite variety of bare-band models (so long as they are differentiable).  Most importantly, by its form
it also explicitly shows that the self-energies are evaluated along the $k_m$ path 
in the case where there is global momentum-dependence in the self-energy.  One might imagine that for the
analysis of a particular experiment one may have reasons to choose one method over another, or perhaps even
a hybrid of the two.  Here we will describe the idea of a Kramers Kronig bare-band fitting (KKBF) as implemented
for our method, its application to other methods is similar.

\begin{figure}[t]
\includegraphics[width=1.0\linewidth]{\figdir/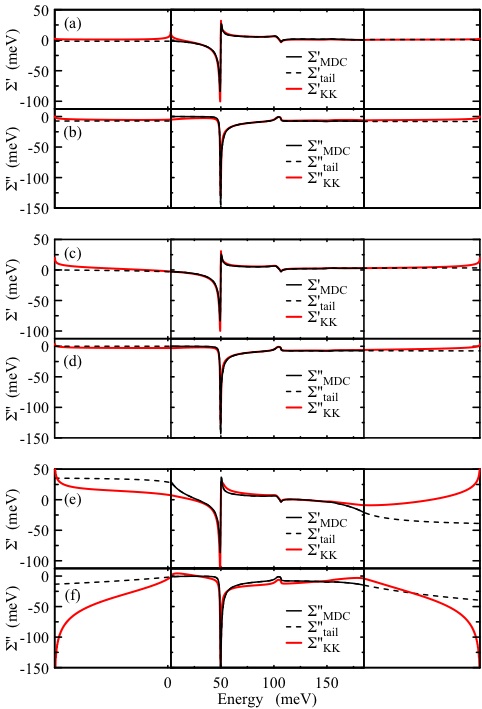}
\vspace{-0.45cm}\caption{(Color online). Self-energies as defined in Fig.\,\ref{fig:kkbf_results}
for the MA$^{(1)}$ Holstein problem with $\Omega\!=\!50$\,meV and $\lambda\!=\!0.15$,
 with extrapolated tails for $\Sigma_{\text{MDC}}$ and the KK transform
shown.  Panels (a) and (b) have the bias used in fitting the tails set
too small, (e) and (f) have the bias set
too large, and (c) and (d) have it set just right.}\label{fig:tails}
\end{figure}

KKBF is a technique whereby a Lorentzian fit is first performed on every slice of constant
energy, $\tilde\omega$, according to Eq.\,\ref{eqn:se_lorentzian}.  The values of $k_m$ and
$\Delta k_m$ from the fits can provide the self-energies for every $(\tilde\omega,k_m)$ point, within the
limitations above, if the bare-band $\varepsilon^b_{k}$ is known.  As an analytic complex function the real
and imaginary parts of the self-energy are Kramers-Kronig related:

\begin{equation}
\Sigma^{\prime\,,\,\prime\prime}_{KK}(k,\omega) = \pm \frac{1}{\pi} \mathcal{P} \int^\infty_{-\infty}
\partial \xi
\frac{\Sigma^{\prime\prime\,,\,\prime}_{MDC}(k,\xi)} {\xi-\omega}. \label{eqn:kk}
\end{equation}
\noindent

It is possible to `fit' the bare-band parameters by choosing them such that 
$\Sigma^{\prime}_{MDC}\!\equiv\tilde\omega\!-\!\varepsilon^b_{k_m}$ and
$\Sigma^{\prime\prime}_{MDC}\!\equiv\!- v^b_{k_m} \Delta k_m$ are
self-consistent with $\Sigma^{\prime}_{KK}$
and $\Sigma^{\prime\prime}_{KK}$.  Since neither the Kramers Kronig relationships (Eq.\,\ref{eqn:kk})
 nor the MDC relationships (Eq.\,\ref{eqn:mdc_params}) are sensitive to a constant offset in both the 
real self-energy and bare-band this is unconstrained by the method
and both $\Sigma^{\prime}$ and $\varepsilon^b_k$
are free.  In our study we have simply made the calculation of differences between
$\Sigma_{KK}$ and $\Sigma_{MDC}$ insensitive to a constant offset, and set the final offset to zero by hand
at the end to allow comparison.

In our implementation of this idea, a simple third order polynomial was used to fit the bare-band with an
initial guess found by fitting MDC peak
maxima.  We then used the Levenberg-Marquardt Algorithm \cite{LMalg}
 as implemented in the mpfit package for IDL \cite{mpfit} to vary band
parameters.
We found that the standard sum-of-squares minimization on the differences between $\Sigma_{KK}$ and 
$\Sigma_{MDC}$ did not perform as well as a concave-down function, as it placed too much
weight on outlying points far away.
In order to evaluate the integrals in Eq.\,\ref{eqn:kk} with a finite
region of data 
tails were extrapolated before a Fourier-based
transform was performed (the tails were then discarded, leaving the
analysis of MDC and KK curves only within the data region).
These tails were extrapolated by fitting an inverse polynomials to each
side of the data, weighing the fit for each side's tail with an exponentially
decaying bias parameter.  A bias parameter of zero would weigh the
entire curve equally, while a large bias parameter would concentrate
only near that data edge.
 
It is possible for problems such as tails, overweighted outliers, and unconstrained offsets to 
compound each other.  An unconstrained constant offset in $\Sigma_{MDC}^{\prime}$ 
and $\varepsilon^b_k$ leads toward
a tendency for a small linear offset in both, which when Kramers-Kronig transformed will distort 
$\Sigma_{KK}^{\prime\prime}$ most visibly near the edges of the data, where it can interfere
with a good fit of the tails.  This in turn can lead to inaccuracies
at these edges,
which if overweighted can distort the bare-band fit itself.  This runaway condition results in a
fit which gets progressively worse through iterations and will never find the correct bare-band.
In practise we found that the tail bias parameter as
well as the the concavity of the function used to process errors must be carefully 
adjusted by hand in order to prevent
this, which can be accomplished simply by looking at whether or not the tail approximation continues to
appear
reasonable through successive iterations.

In Fig.\,\ref{fig:tails} we show some typical examples of
how the tail bias parameter can affect the fitting, each pair of panels represents the final "solution" of the
entire band minimization problem using a given tail bias.  Plots like these form the guide to
be used when adjusting the bias
parameters by hand while looking for the best solution.
In panels (a) and (b) the tail bias is too small, and so the found tail is the best
approximation which fits the whole curve.  In panel (b) this causes a discontinuity for the low-energy
tail right at the boundary, which in turn causes a cusp in the KK transform visible in (a).
Despite this, the overall fit is not too
bad with reasonable general agreement between MDC and KK self-energies - 
meaning the found bare-band is likely close to the real solution.
In panels (c) and (d) the tail bias is good, which results with a realistic fit at all boundaries and a good
agreement between MDC and KK self-energies giving confidence that the found bare-band is accurate.
In panels (e) and (f) the tail bias is too strong, which results in a tail fit depending too much on the data
right at the edges.  This results in a KK transform which is poor enough to thrown off the band fitting entirely
resulting in a found bare-band which is likely not close to the true band, shown by generally poor agreement
between MDC and KK self-energies.

\section{acknowledgements}

We gratefully acknowledge S. Johnston, T.P. Devereaux, F. Marsiglio
I.S. Elfimov, B. Lau, and G.A. Sawatzky for many useful discussions. This
work was supported by the Killam Program (A.D.), Alfred P. Sloan Foundation (A.D.), CRC Program
(A.D.),
Steacie NSERC Fellowship Program (A.D.), 
NSERC, 
CFI, CIFAR Quantum Materials and Nanoelectronics Programs, and BCSI.

\bibliography{sf_tour}

\end{document}